# Large Size GEM for Super Bigbite Spectrometer (SBS) Polarimeter for Hall A 12 GeV program at JLab

Kondo Gnanvo[1*], Nilanga Liyanage[1], Vladimir Nelyubin[1], Kiadtisak Saenboonruang[1†], Seth Sacher[1],

Bogdan Wojtsekhowski[2]

[1]*University of Virginia, Department of Physics, Charlottesville, VA 22903, USA*

[2]*Thomas Jefferson National Accelerator Facility, Newport News, VA 23606, USA*

**Abstract**

We report on the R&D effort in the design and construction of a large size Gas Electron Multiplier (GEM) for the Proton Polarimeter Back Tracker (BT) of the Super Bigbite Spectrometer (SBS) in Hall A at Thomas Jefferson National Laboratory (JLab). The SBS BT GEM trackers consist of two sets of five large GEM chambers of size $60 \times 200$ cm$^2$. The GEM chamber is a vertical stack of four GEM modules, each with an active area of $60 \times 50$ cm$^2$. We have built and tested several prototypes and the construction of GEM modules for SBS BT is ongoing. We describe in this paper the design and construction of the GEM module prototype as well as the preliminary results on performance from tests carried out in our detector lab and during test beam at Fermi National Laboratory (Fermilab).

## 1. Introduction

In 2014, Jefferson Lab [1] will complete the 12 GeV upgrade of its Continuous Electron Beam Accelerator Facility, (CEBAF) which will allow outstanding study of nucleon structure in terms of form factors (FFs), transverse momentum distributions (TMDs), generalized parton distributions (GPDs) and structure functions in the valence quark region. CEBAF will be capable of delivering longitudinally polarized electron beams at an intensity of up to 85 µA, which corresponds in some experiments to luminosities of $10^{39}$ electron/s-nucleon/cm$^2$. The SBS apparatus [2] in Experimental Hall A combines a high performance and large acceptance detector package (with an active area of ~1 m$^2$) in an open geometry configuration with a dipole magnet located close to the target to provide a relatively large solid angle up to 75 msr for experiments to run at the largest luminosity possible. The SBS detector package can be arranged in specialized configurations to meet the needs of a number of highly-ranked experiments for the 12 GeV program, namely $G_E^n$, $G_E^p$ and $G_M^n$ [2].

---

[*] Corresponding author: Tel.: +1 434-9246796, Fax, +1 434-924-4576, e-mail: kgnanvo@virginia.edu

[†] Currently, Lecturer at Department of Applied Radiation and Isotopes, Faculty of Science, Kasetsart University, Bangkok, Thailand 10900

The SBS, however, requires very high counting rate detectors to cope with a large background rate. Background rates as high as 400 kHz / cm$^2$ [2] are expected in its Front Tracker (FT), the closest to the target, for experiments such as $G_E^p$ for such high luminosity. Moreover, the small field integral (2-3 Tesla-m) of the magnet means that one needs an angular resolution of 0.3 mrad [2] to achieve a momentum resolution below 1% with the small bend angle of incoming protons at a momentum of 8 GeV/c. With the FT located at about 70 cm from the target, in order to achieve this level of precision measurement, detectors with spatial resolution of the order of 70 μm must be used. GEM technology [3] has been adopted for the tracking systems of SBS to meet these challenges. The GEM technology permits large area detectors with high counting rate capabilities, exceeding to 2.5 MHz/cm$^2$ [4], and excellent spatial resolution of around 70 μm [5]. The detector group at the University of Virginia (UVa) is leading the design and fabrication of SBS BT GEM chambers for the SBS Polarimeter. In this paper, we report on the effort to design, build and characterize the SBS BT GEM modules.

## 2. Large area GEM Trackers for SBS Polarimetry

### 2.1. Layout of the SBS proton arm for GEp (5) experiment.

The most demanding SBS experiment, GEp (5) [6], is an ambitious physics program to measure the ratio of electric and magnetic elastic FFs of the proton at a 4-momentum transfer (Q$^2$) up to 12 GeV$^2$. For this experiment, the SBS will be configured to detect the recoil proton and measure its polarization. As such, the SBS requires three GEM trackers, the FT, the Second Tracker (ST) and the Third Tracker (TT).

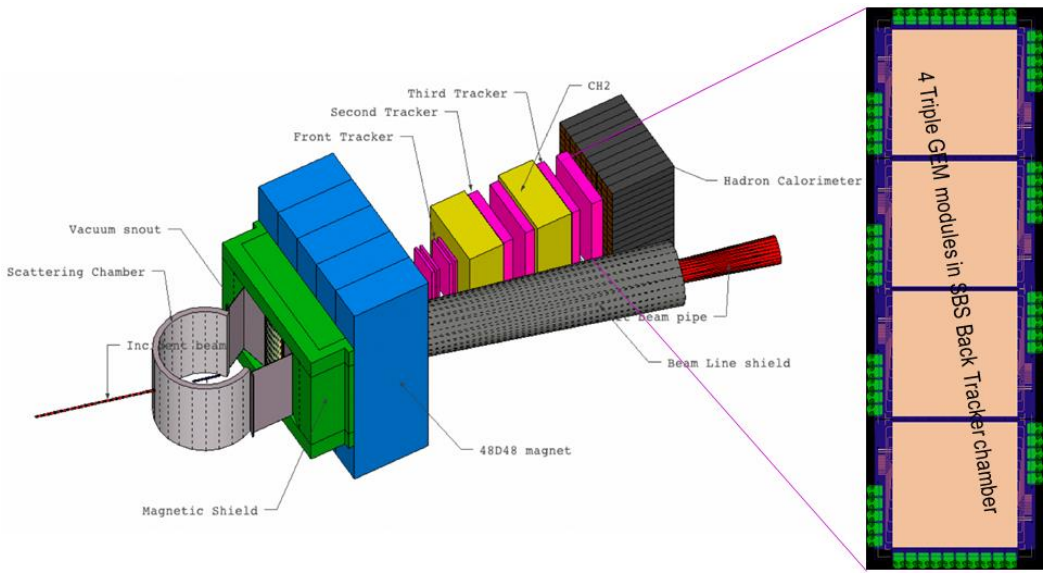

Figure 1: (Left) layout of Super Bigbite Spectrometer for the GEp (5) experiment; (right) layout of SBS BT GEM chamber. The chamber is composed of 4 GEM modules in a vertical stack.

The ST and TT constitute the Back Tracker (BT). Figure 1 shows the layout of the proton arm of the SBS for the GEp (5) experiment and one SBS BT GEM chamber. The FT is used to measure, with high precision and in a large background environment, the momentum and direction of the recoil protons from the electrons scattering off the target. It is made of six large (40 × 150 cm$^2$) GEM chambers. Each FT chamber is a vertical stack of three GEM modules, with an active area of 40 × 50 cm$^2$. The FT GEM modules are being designed and fabricated by INFN-Roma and INFN-Catania in Italy [7]. The ST and TT are combined with two CH$_2$ analyzer blocks for proton polarimetry. Proton polarimetry technique allows the determination of the proton FF ratio by the method of polarization transfer from the measurement of the ratio of the longitudinal and transverse components of the proton spin polarization. The transverse polarization of the recoil proton is measured from the azimuthal asymmetry after rescattering in one of the two CH$_2$ analyzer blocks. The polar and azimuthal angles of the rescattered proton are measured in the tracker downstream of the analyzer block, with respect to the angles measured by the upstream tracker [2]. Each SBS BT GEM chamber is a vertical stack of four GEM modules, each with an active area of 60 × 50 cm$^2$. A sketch of a SBS BT chamber is shown at the right hand side of Figure 1. A total of 40 modules (of 60 × 50 cm$^2$) are required for 10 SBS BT GEM chambers (60 × 200 cm$^2$). The R&D, prototyping and the construction of the SBS BT GEM modules is done at UVa.

## 2.2. Design of the SBS Back Tracker GEM module

The first two SBS BT GEM prototypes have an active area of 50 × 50 cm$^2$ whereas in the final design for the SBS, the SBS BT GEM modules have a larger active area of 50 × 60 cm$^2$ in order to improve further the acceptance of the double polarimeter. The results reported in the present paper mainly concerns the design and extensive tests carried out with the 50 × 50 cm$^2$ SBS BT GEM prototype but remain equally valid for larger production modules for BT chamber. The picture on the left of Figure 2 shows the first SBS BT GEM prototype. The sketch on the right of Figure 2 is a cross sectional view of the SBS BT GEM module. The SBS BT GEM module design is based on R&D work previously done on GEM detectors for the COMPASS experiment [4, 5] at CERN and for the Muon Tomography project [8] at the Florida Institute of Technology (FIT).

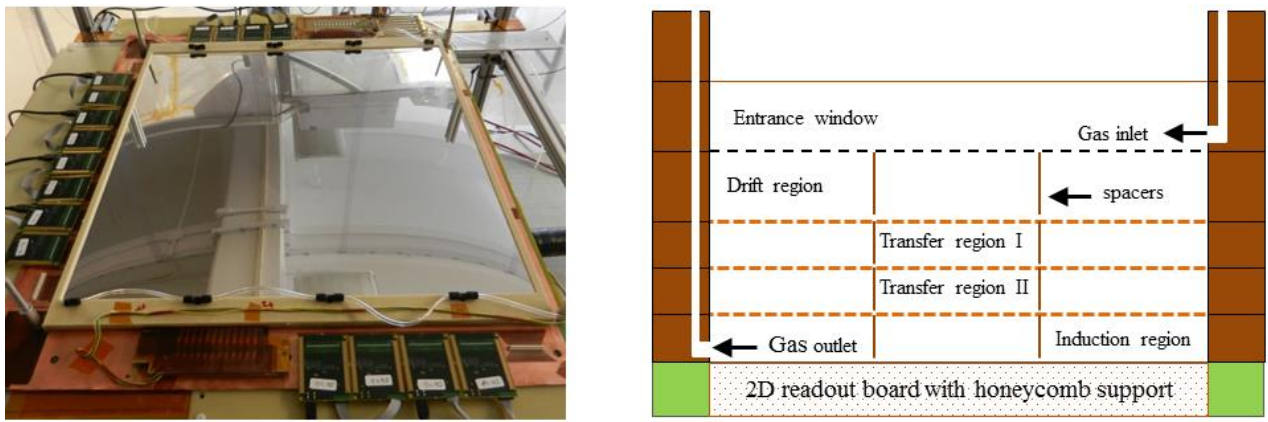

Figure 2: (Left) SBS BT GEM prototype; (right) cross section of the SBS BT GEM module.

It consists of a 3 mm ionization gap between the drift cathode and the top GEM foil in the stack of three GEM foils. The GEM foils are separated from each other by a 2 mm transfer region and provide three stages of gas avalanche amplification of the primary ionization. This detector configuration with 3 GEM foils for the electron amplification is referred to as Triple-GEM detector. The drift of the electron cloud from the last GEM foil to the flexible readout board, located 2 mm below the GEM stack, induces signals in the 2D Cartesian readout strip layers. The design is optimized to operate at a rate as high as 1 MHz / $cm^2$ while providing excellent spatial resolution of 70 μm with a stable and non-flammable $Ar/CO_2$ (70/30) gas mixture [4,5]. The active area of the SBS BT GEM module is, however, more than three times larger than COMPASS or FIT Triple-GEM detectors. Because of their large size, the foils used for the SBS BT GEM modules are fabricated at the CERN Printed Circuit Board (PCB) workshop using the single mask technique [9] developed at CERN for large area GEM.

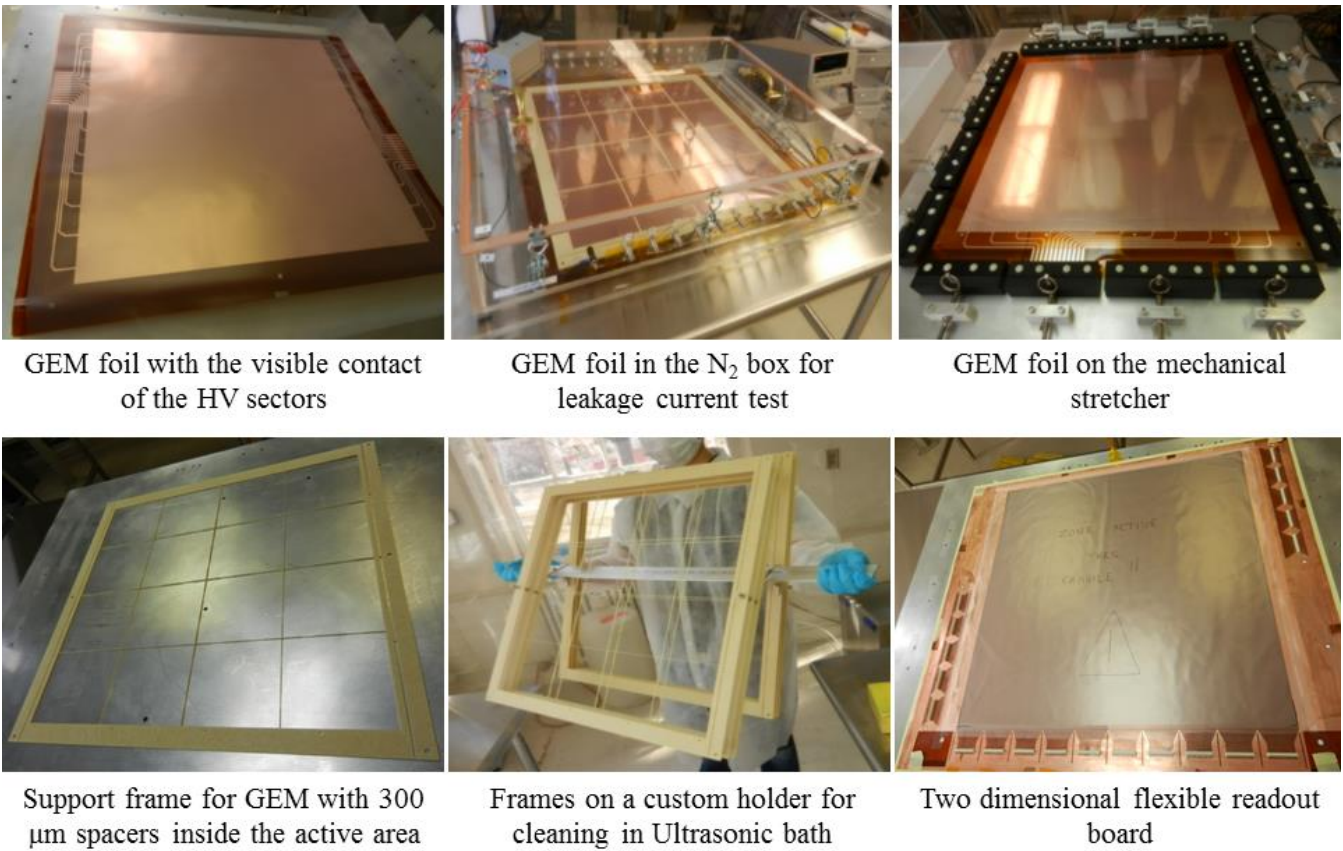

Figure 3: Components of SBS BT GEM module and a few clean room assembly steps.

A GEM foil for a SBS BT GEM prototype is shown in the top left picture of Figure 3. The 5 μm thick Cu electrode on the top side of each GEM foil is segmented into 24 high voltage (HV) sectors of roughly 100 $cm^2$, (30 HV sectors for the $50 \times 60$ $cm^2$ production modules). The GEM foils and the drift cathode are stretched and glued onto supporting frames made of the fiberglass-reinforced epoxy material Permaglas[1]. The frames have a grid of 300 μm wide spacers to maintain a uniform gap (therefore a uniform electric field) between the layers after assembly. Pictures of the frames are shown at the bottom left and center of Figure 3. The two horizontal sides of the supporting frames are narrow, with 8 mm widths, to minimize the dead-to-active area ratio when four modules are combined into one SBS BT GEM chamber as shown in the right picture of Figure 1. The wider (30 mm) vertical sides of these frames provide room for gas distribution holes and high voltage lines, as well as extra support for mechanical tension of the GEM foils. A GEM foil with the copper stripped off of one side is used as the cathode foil to allow for a simple but efficient gas flow throughout the entire sensitive volume of the detector. The 2D Cartesian readout board, based on the COMPASS GEM readout design, and fabricated using the same copper-clad Kapton material as for the GEM foils, consists of two sets perpendicular strips (top and bottom), with 400 μm pitch. The top strips are 80 μm wide and are isolated from the bottom by 50 μm thick Kapton layers, while the bottom strips are 350 μm wide. The width and the gap between top and bottom strips have been chosen to ensure an equal sharing of charge. The readout board is shown in the bottom right picture of Figure 3. The drift cathode and the readout board for the SBS BT GEM modules are also fabricated at the PCB workshop at CERN.

## 3. Module Construction Procedure

### 3.1. Support frame preparation

The support frames are first polished with sand-paper to remove any protruding glass fibers ends, and then cleaned in an ultrasonic bath filled with demineralized water for 15 minutes. The frames are then dried in air in the clean-room for 3 to 4 days. An electrical test of the frames is performed by placing them in between two metallic plates and applying a voltage of 4 kV between the plates. Finally, the parts of the frame exposed to the sensitive volume of the chamber are coated by applying, with a hand brush, a thin layer of polyurethane, Nuvovern LW + Hardener[2] to prevent surface irregularities, residual fiber ends or sharp edges.

### 3.2. High voltage test of the GEM foils

Each of the 24 high voltage (HV) sectors of a GEM foil is tested in the dry Nitrogen ($N_2$) box by applying a voltage ΔV = 550V across the foil through the sector top electrode and the bottom electrode and measuring the leakage current. The criterion for an acceptable foil is a current of less than 5 nA under HV for 2 min. During the assembly, the test is performed first on the bare foil and then on the framed foil. In both cases the foil is placed in the box with an open flow of $N_2$ (picture in top middle of Figure 3). When the chamber is completed, the test is repeated on each individual sector with $N_2$ gas flowing in the chamber. The distribution of the combined leakage current data from all 72 HV sectors of the foils used for the prototype is shown in Figure 4, with respectively from left to right, the measurement on the bare foils, on the framed foils and after the final assembly in the chamber. The mean value is well below the 5 nA/sector design goal.

---

[1] from Permali, distributed by Resarm Engineering Plastics in Belgium, http://www.resarm.com

[2] Polyurethane distributed by Walter Mäder AG, http://www.maederlacke.ch

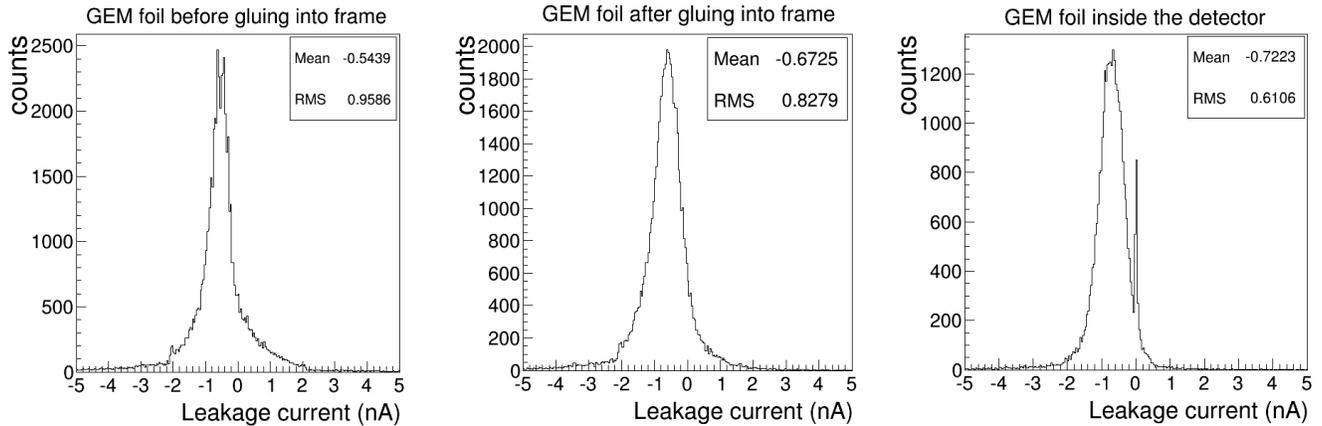

Figure 4: Distribution of the leakage current over 72 HV sectors of the GEM foil before framing (left), after framing (middle) and inside the assembled chamber (right).

### 3.3. Stretching and gluing of the GEM foils

For the stretching of GEM foils, we use a custom-made mechanical stretcher, based on the original design by Alfonsi *et al.* [10], as shown in the picture on the top right of Figure 3. The base of the stretcher is a 0.5 inch thick aluminum template with alignment holes to match the holes in each of the support frames and GEM foils. The GEM foil is positioned and aligned on the aluminum template with dowel pins and clamped into the stretching jaws on all four sides. At two adjacent sides, the jaws are equipped with load cells connected to digital displays for monitoring the tension applied to the foil. A uniform tension of 0.5 kg/cm, corresponding to a pressure of 1 MPa, is applied along the foil. The spacer frame is glued onto the stretched foil using a resin epoxy (Araldite AY103 + HD991 Hardener)[3]. This glue is convenient for its excellent electrical behavior, well-studied aging and low outgassing properties. The framed foil is validated with a high voltage test of each sector as indicated above.

### 3.4. Final assembly

The base of the GEM module is formed by gluing the 2D readout onto the honeycomb support placed on an assembly support template and aligned with dowel pins. Then the three framed GEM foils, the framed cathode frame and the gas window are respectively glued onto the base. Finally, the gas connectors and tubes are mounted on the chamber to close it from the outside environment. Once these procedures are completed, the GEM module is removed from the clean room and a thin layer of Dow Corning 1-2577 RTV Coating[4] is applied to all four sides to prevent further gas leaks. The 72 HV sectors of the 3 GEM foils in the chamber are once again tested by applying a voltage $\Delta V = 550$ V across the GEM foils with $N_2$ gas flowing inside the chamber before the HV board is mounted. Figure 5 shows a schematic and a picture of the HV board, with its main component, a custom-made ceramic resistive divider available at CERN PCB workshop. The resistive divider consists of a resistor chain with 8 metallic contacts to distribute the voltage across the 6 GEM electrodes, the drift cathode and the ground from a single input HV power supply.

---

[3] http://commerce.sage.com/bag/upload/site/default/FT/Fran%C3%A7ais/Araldite/Araldite%20AY%20103-1%20+%20HY991.pdf

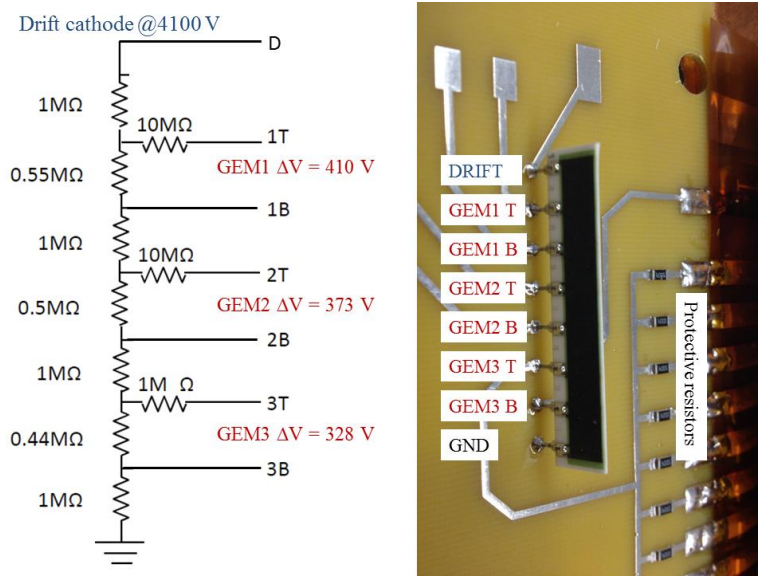

Figure 5: (left) Schematic of the resistive divider with the voltage across the GEMs for HV = 4.1 kV on the HV board; (right) HV board with the resistive divider and the protective resistors mounted on SBS BT GEM prototype

The HV board also hosts 10 MΩ protection resistors connected to each individual HV sector of GEM foils 1 and 2 and 1 MΩ resistors connected to the HV sectors of GEM foil 3. These resistors are used to limit the damage due to discharge caused by a spark during the operation. Once the HV board is installed, a voltage scan, up to 4.2 kV across the resistive divider, corresponding to an average voltage drop $\Delta V = 382$ V across a GEM foil, is performed with $N_2$ flowing inside the detector. The chamber is kept at the high voltage 4.2 kV for about 24 hours to burn any residual dust inside. Finally, after this last check, the module is filled with an $Ar/CO_2$ (70/30) gas mixture for tests and characterization.

## 4. Performances of SBS BT GEM prototypes

---

[4] http://www.dowcorning.com/applications/search/default.aspx?R=263EN

Two 50 × 50 cm$^2$ SBS BT GEM prototypes were tested with cosmic and radioactive sources as well as in the high energy hadron beam at the Fermilab Test Beam Facility (FTBF) in October 2013. The performance parameters of the detectors such as the gain uniformity, efficiency and charge sharing of the readout board have been studied. The chamber typically operates with an Ar/CO$_2$ (70/30) gas mixture for the tests with the applied high voltage in the range of 4 kV to 4.2 kV on the resistive divider, corresponding to an average voltage ΔV range of 364 to 382 V across the GEM foils. A front end (FE) electronic readout system, the Scalable Readout System (SRS) [11], based on the APV25 readout chip and developed by the RD51 collaboration at CERN [12] was used to read out the SBS BT GEM prototypes. The SRS was run with DATE [13] for the data acquisition and AMORE [13] for the monitoring and data analysis. These two software packages have been developed by the ALICE experiment at CERN. A systematic pedestal run with the APV25 readout electronic connected to the detector is performed before the detector characterization with cosmics, radioactive source or test beam.

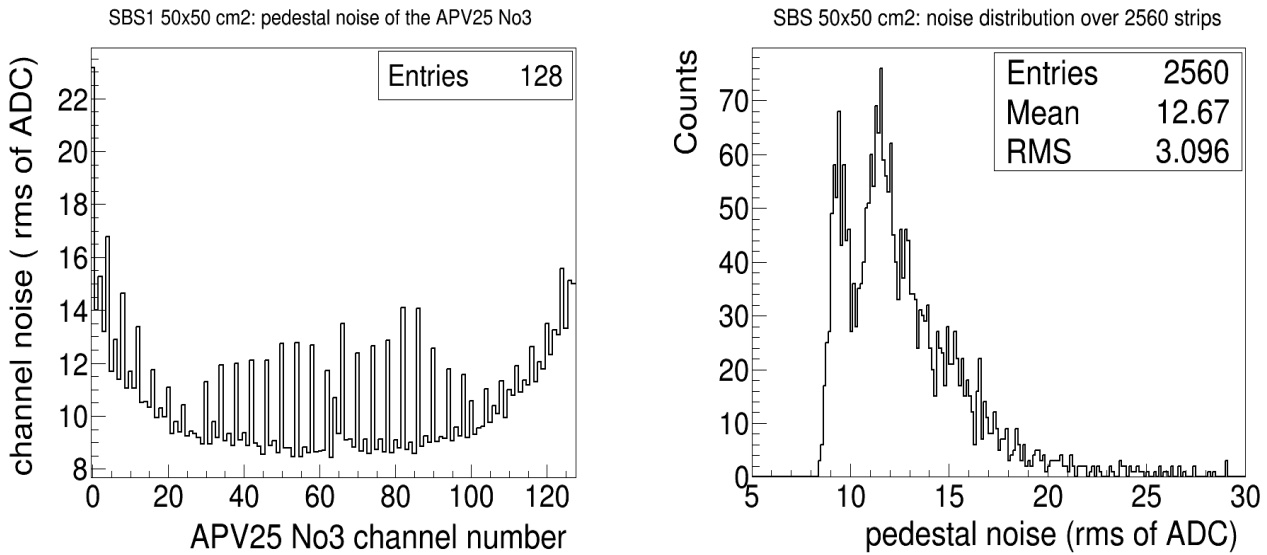

Figure 6: (Left) Pedestal noise i.e. rms for each channels of one APV25 card connected to SBS BT GEM prototype; (right) Distribution of the pedestal rms for all 2560 APV25 channels (20 cards) connected to the prototype.

During the pedestal run, the voltage ΔV across the GEM foils is set below 200 V to prevent cosmic particles from generating signals in the detector. The mean and the rms of the distribution of ADC (Analog-to-Digital Converter) counts representing respectively the pedestal offset and the pedestal noise is then compiled for every APV25 channel and saved into a ROOT file. The left plot in Figure 6 shows the pedestal rms in ADC counts for each of the 128 channels of one APV25 FE connected to the SBS BT GEM prototype. One ADC count is equal to approximately 232 e$^-$ which is equivalent to 0.03712 fC. The rms varies from channel to channel with a minimum around 7 ADC counts and a maximum around 22 ADC counts. A similar pattern is seen for all the APV25 FE and is intrinsic to the internal routing of the channels inside the APV25 FE card. The total pedestal noise is a convolution between the intrinsic noise of the APV25 and the capacitance noise from the detector readout strip. The distribution of the pedestal rms for all 2560 APV25 channels connected to the SBS BT GEM prototype is shown in the right plot of Figure 6. This distribution shows two peaks at respectively, 9 and 12 ADC counts. These peaks correspond to the contributions to the pedestal noise due to the capacitance noise of the narrow top layer strips and the wide bottom layer strips of the prototype readout board as described in section 2.2.

For the characterization of the SBS BT GEM prototype, zero suppression is performed, after pedestal subtraction and common mode correction, by selecting only the strips above a certain threshold defined by (n × $\sigma_{strip}$) with $\sigma_{strip}$ being the rms of pedestal distribution for each strip described above in the example on the left plot of Figure 6.

### 4.1. SBS BT GEM response and gain uniformity

A large set of cosmic data, with over $3 \times 10^6$ events from a weeklong run, was used to study the gain uniformity over the entire area of the chamber. The left plot in Figure 7 shows the 2D distribution of cluster position (x, y). A cluster is defined as a collection of consecutive strips with signals above zero suppression. The lack of events at the edges of the chamber is due to the limited size of the trigger scintillators used for the data acquisition.

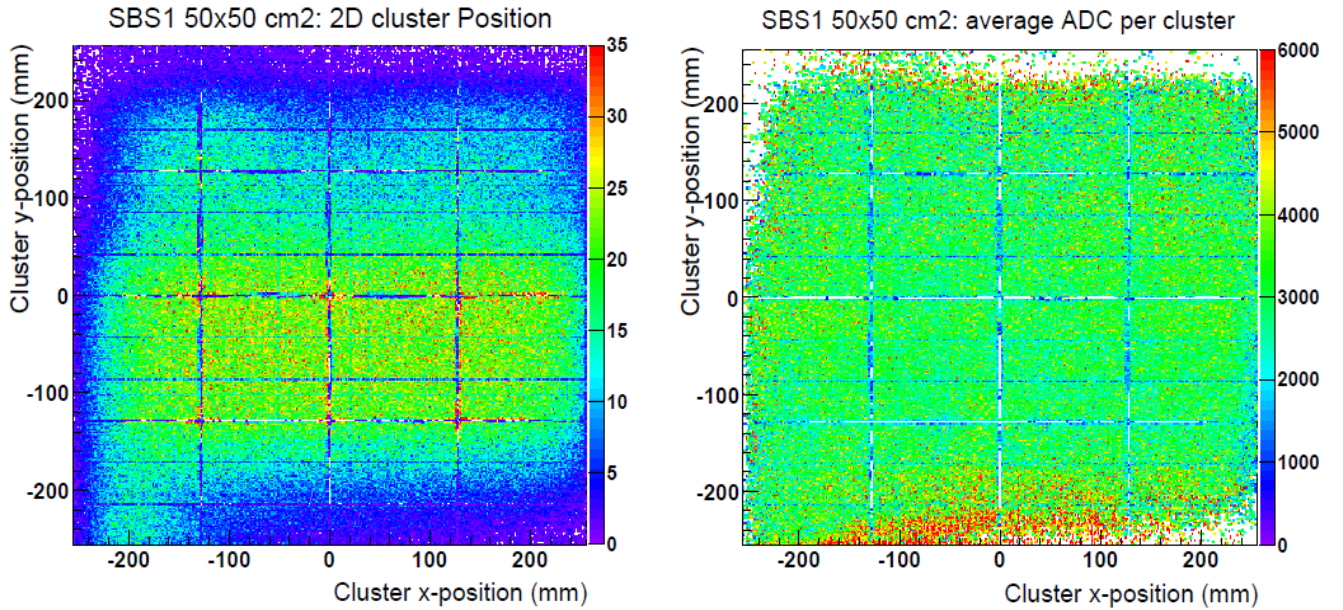

Figure 7: (left) 2D map of cluster position from cosmic data run; (right) 2D spatial distribution of average cluster charge (in ADC counts) with large cosmic data run: relative gain uniformity measurement.

The right plot in Figure 7 shows the 2D spatial uniformity of average gain (in ADC counts) of the SBS BT GEM prototype. The average gain as a function of position was calculated by filling in each bin of a 2D (x versus y) histogram with the total ADC values of all strips of the clusters with position coordinates falling into that bin over the entire run. At the end of the run, the ADC sum in each bin is divided by its count (the number of bin entries) to obtain the average ADC for that bin. The effect of the 300 μm wide spacers of the GEM frames is clearly shown here on the two plots and represents a dead area with an approximate width of 2 mm. The distinctive horizontal lines are due to additional efficiency drop caused by the HV sector boundaries on the GEM foils.

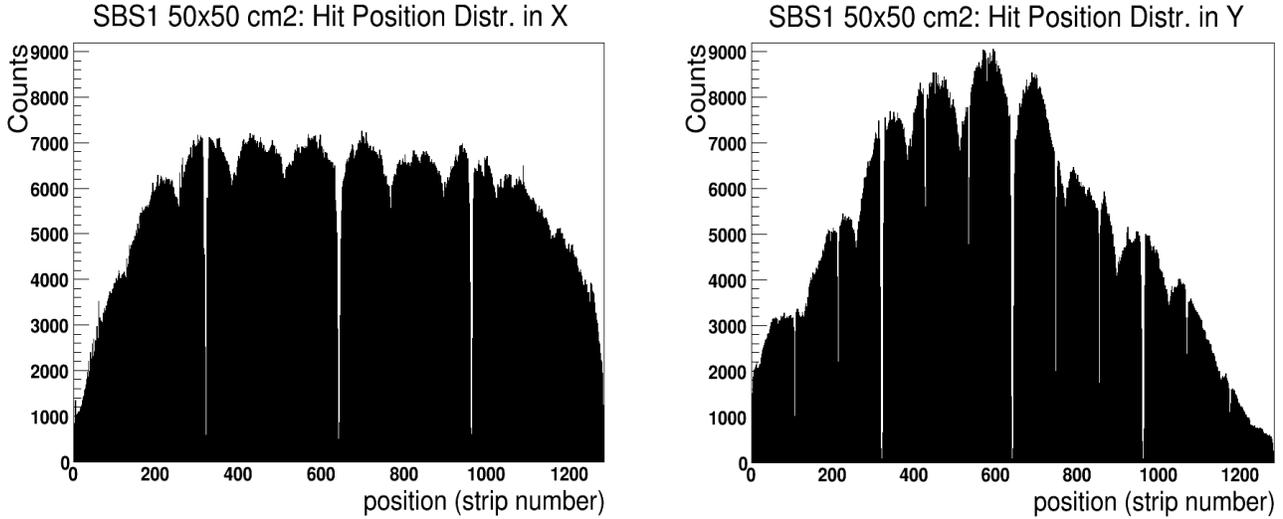

Figure 8: Distribution of the hits from cosmic data across the 1280 strips in x (left) and y (right); drops in efficiency caused by the spacers and HV sectors boundaries are clearly visible.

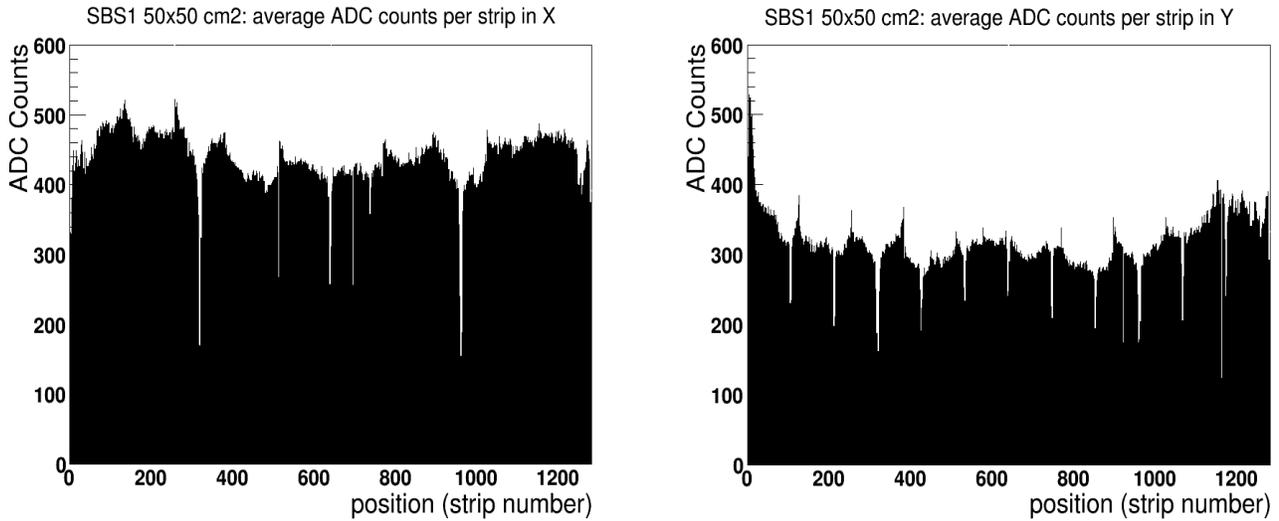

Figure 9: Distribution of average ADC counts per strip on the 1280 strips in x (*left*) and y (*right*) from cosmic data

Left Figure 8 shows the cumulative strip occupancy of both x-plane and y-plane (1280 strips each) in the prototype from the same large cosmic data sample. The expected drops in the count distributions caused by the presence of the spacers in x-plane and y-plane as well as the HV sectors boundaries for y-plane are once again clearly shown. The wave-like structure seen on the two plots is the manifestation of the non-uniformity of the pedestal noise of the APV25 FE card described earlier in section 4, (see left plot in Figure 6). The periodicity of this pattern on the strip occupancy plots reflects the number of cards connected to the detector under test. The average gain (in ADC counts) per strip is shown in Figure 9. Each bin of the histograms represents the relative gain of the strip calculated using the same method as for the 2D relative gain uniformity map of Figure 7. The distribution shows good uniformity of the relative gain with less than 10 % variation in the x- direction and 15% in the y-direction. It should be noted that these results have been obtained without correcting non uniformities in the electronic response of the APV25 readout chip; this means that the effective gain variation of the detector could be even lower than what we observed.

**4.2. Charge sharing of the 2D readout strips layer**

The two SBS BT GEM prototypes were tested at FTBF in October 2013 as part of the slice sector test campaign by RD6-FLYSUB[5] Consortium. The prototypes were installed on the FTBF MT6.2b stand in a setup that included 10 other GEM detectors with various sizes ranging from $10 \times 10$ cm$^2$ small Triple-GEMs to 1 m long trapezoidal GEM. The first SBS BT GEM prototype was mounted on a moving table during the test beam. FTBF provides a 120 GeV primary proton beam as well as secondary hadron beam consisting in a mixture of kaons, pions, protons at energy ranging from 20 GeV to 32 GeV.

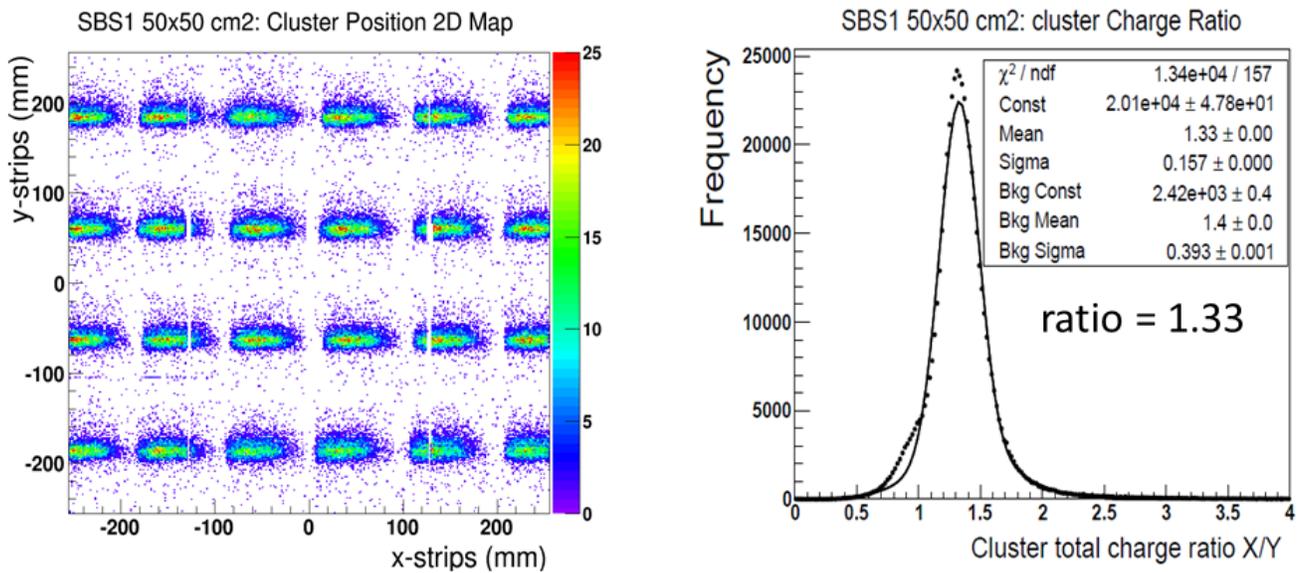

Figure 10: (Left) distribution of 32 GeV hadron beam reconstructed from 24 positions; (right) x vs. y charge sharing

---

The beam has a cycle of 1 min and 4 second spill of up to $2 \times 10^4$ particles. The facility also provided the external trigger for our readout electronics with the coincidence signal from 3 Photomultiplier Tube (PMT)-scintillators located upstream and downstream of our test beam setup. The performances of the prototypes were studied with the data taken during the test beam campaign. The left plot in Figure 10 shows data collected from a 32 GeV hadron beam at 24 different positions on the first SBS BT GEM prototype. Each spot on the plot represents the 2D profile of the beam with an approximate size of 8 cm × 2 cm at one given location. The right plot in Figure 10 shows the charge sharing ratio equal to 1.33 between the x (top strips) and y (bottom strips) for this prototype. The non-equal sharing was caused by an issue related to the quality of the 2D readout boards used for the BT prototypes. Ideally, the Kapton material that insulates the top strips from the bottom strips has a thickness of 50 μm and a width of 80 μm (which is same as width of the top strips). In reality, because of the chemical etching process during fabrication, the Kapton always displays a wedge-like shape with a larger width at its base than at the top, covering part of the exposed area of bottom strips. The left picture of Figure 11 shows a microscopic view of one of a readout board from the same batch as the ones used for the SBS prototypes. As the picture indicates, the Kapton layer is as wide as 160 μm (black arrows) at its base; this is twice the expected ideal width of 80 μm (white arrows). The excess Kapton material covers a portion of the bottom readout board reducing the total amount of charge collected by the bottom strips and therefore resulting in a non-equal charge sharing between the top and bottom strips.

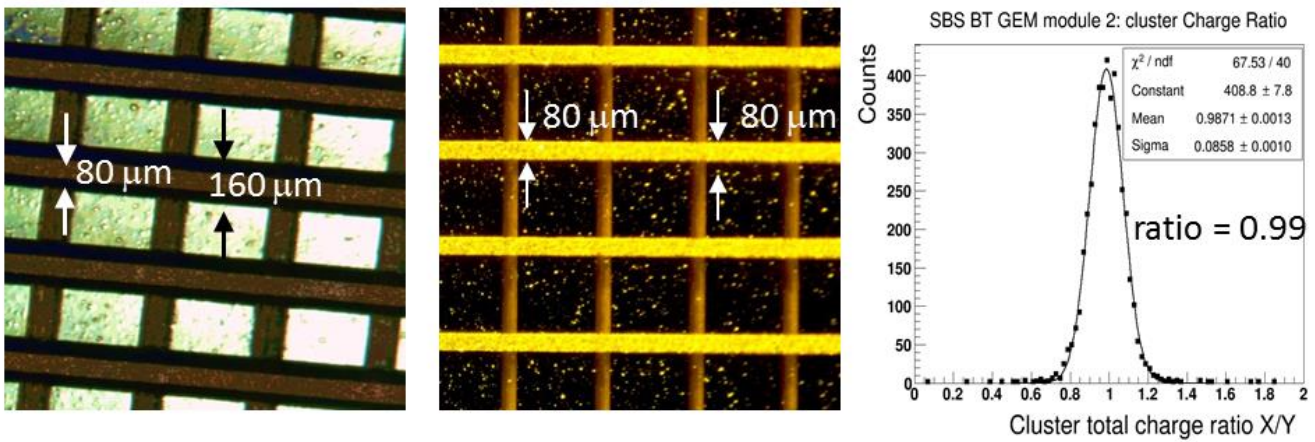

Figure 11: (Left) Poor quality readout board used for SBS BT GEM prototype module; (center) improved readout board quality used for SBS BT GEM production module; (right) equal charge sharing from a production module with improved quality.

Moreover, there is further degradation of charge sharing when the GEM detector is operated in a high rate environment due to the charging of the dielectric Kapton, which increases the shielding of the bottom strip layer. Since this issue was observed, the production procedures for SBS readout boards at the CERN PCB workshop were improved to ensure the required quality. Furthermore, a quality assurance process, using inspection with a high power microscope, was setup at the UVa detector lab to ensure that the Kapton width at the base is smaller than 100 µm for all readout boards. The picture at the center of Figure 11 shows a new readout board of the required quality used for a newly built SBS BT GEM module; the plot on the right of Figure 11 shows the charge sharing ratio of approximately 1 obtained from the new module.

### 4.3. Efficiency and cluster size

The efficiency of the detector is defined as the number of events with at least one cluster in each of both readout layers (x-strips and y-strips) of the detector, divided by the number of triggered events. The left plot in Figure 12 shows the efficiency curves for 3 different thresholds from a HV scan of the SBS BT GEM prototype. An efficiency plateau above 96% is reached at 4100 V, equivalent to an average voltage $\Delta V = 373V$ across the GEM foils, for a $4 \times \sigma_{strip}$ threshold. The right plot in Figure 12 shows the cluster size, defined as the average number of the hits in the cluster, versus the average voltage $\Delta V$ across the GEM for the x and y planes of the prototype. The cluster size is smaller for the y-strips because of the same readout quality issue discussed in the previous paragraph. These results are consistent with what has been observed with similar GEM detectors before [7].

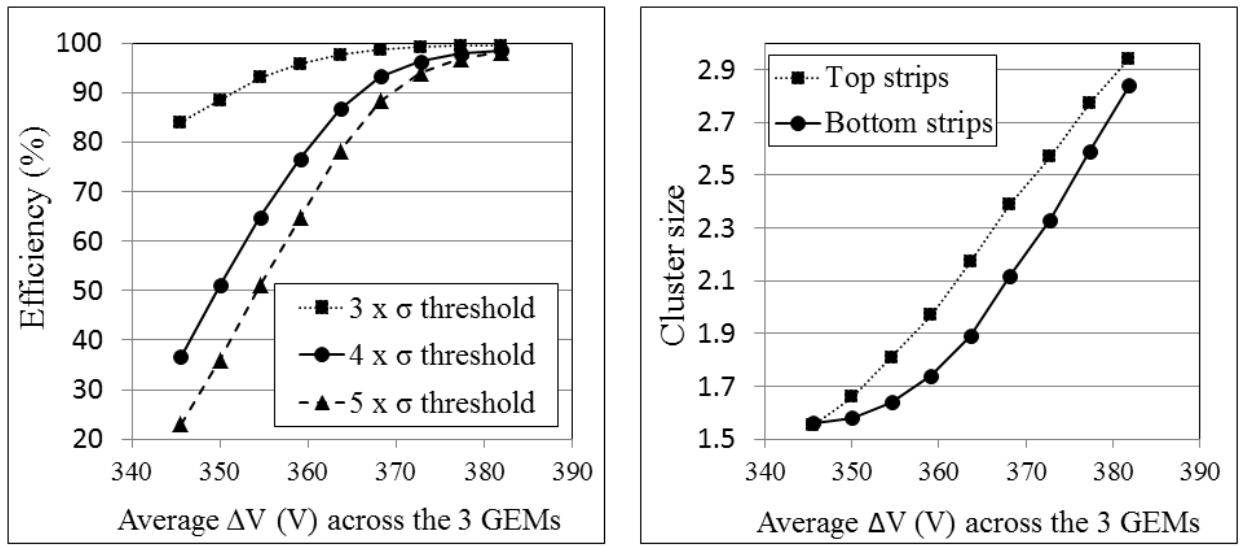

Figure 12: (Left) Efficiency vs. average voltage ($\Delta V$) across the 3 GEM foils for various zero suppression thresholds; (right) Cluster size vs. $\Delta V$ across the 3 GEM foils of both x-and y-strips at $5 \times \sigma$ threshold.

### 4.4. Spatial resolution studies

The spatial resolution of the SBS prototype was also studied with FTBF test beam data. A detailed description of the setup and beam condition as well as the analysis and results will be published soon in a forthcoming paper and only a short summary of the resolution calculation is given here. The setup used during the FTBF test beam for the spatial resolution studies comprised of three small ($10 \times 10$ cm$^2$) standard CERN Triple-GEMs used as trackers.

Two of the small Triple-GEMs were located upstream of the SBS prototype with respect to the beam and the third one downstream. A simple least squared linear fit algorithm was used to fit the tracks. We used the geometric mean method described in [14] to calculate the spatial resolution in the x and y directions of the SBS BT GEM prototype. With this method, the resolution is obtained using the geometrical mean $\sigma = \sqrt{(\sigma_{incl} \times \sigma_{excl})}$ of the width of the track residual distribution of respectively inclusive and exclusive fitted tracks. For the inclusive fit, the data from both probed detectors (the SBS BT GEM prototype) and tracking detectors (small Triple-GEM trackers) are used to get the fitted track whereas in the case of the exclusive fit, only the data from the tracking detectors are used for the fitted track. The SBS BT GEM prototype has in principle the same characteristics as the small Triple-GEM trackers, and thus the FTBF setup provides the ideal configuration for an accurate estimation of the spatial resolution using the geometric mean method [14]. Figure 13 shows the inclusive and exclusive residuals for the x-coordinate, with $\sigma_x$ (excl.) ≈ 69 µm and $\sigma_x$ (incl.) ≈ 51 µm respectively, and for y-coordinate, with $\sigma_y$ (excl.) ≈ 77 µm and $\sigma_y$ (incl.) ≈ 58 µm respectively, at the P1 location on the detector.

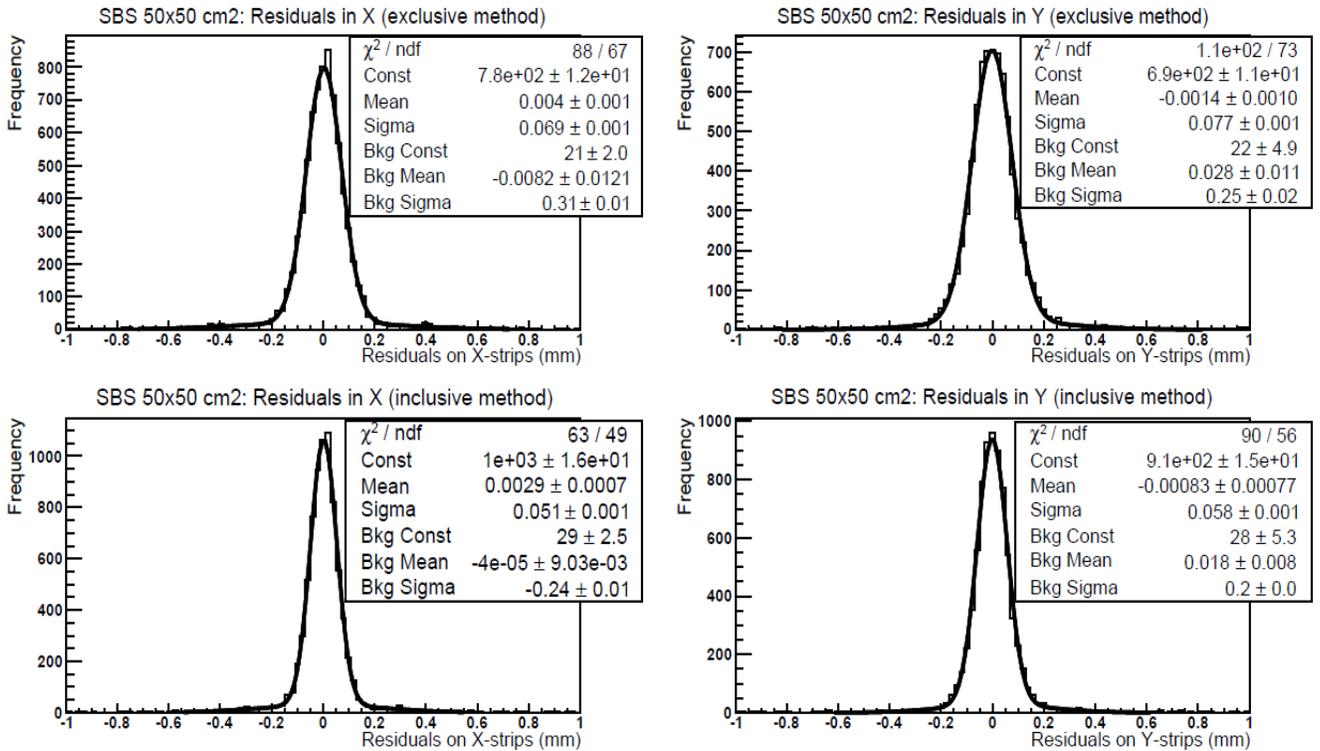

Figure 13 Track residual distribution from the exclusive (top) and inclusive methods (bottom) for x-coordinates (left) and y-coordinates (right) for the SBS BT GEM prototype.

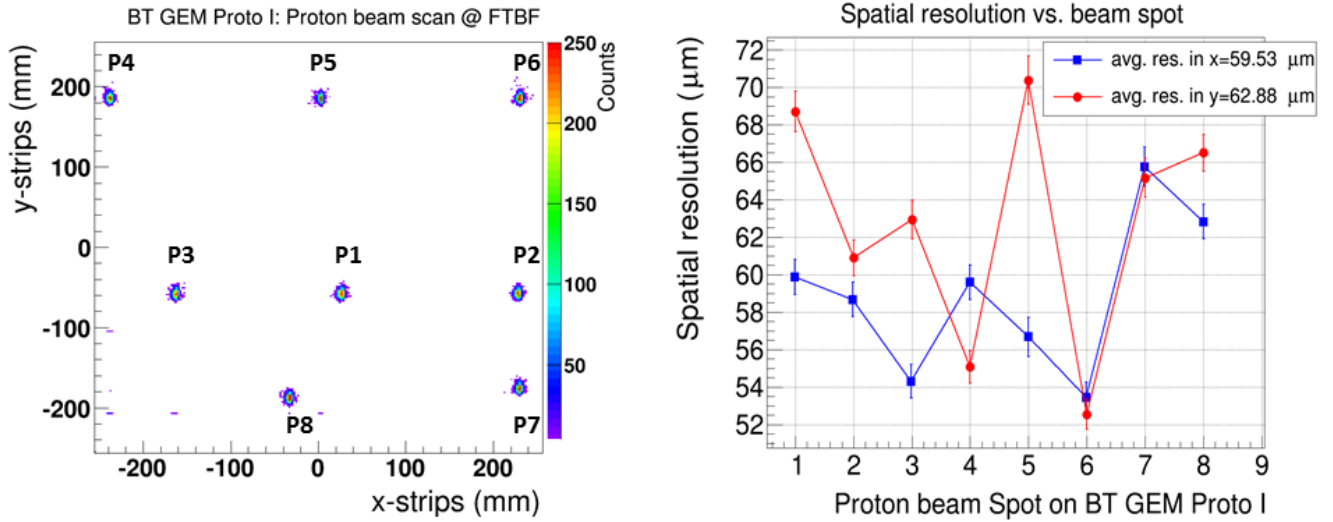

Figure 14: (Left) 2D distribution of 120 GeV proton beam reconstructed from the eight positions scan during FTBF test beam; (right) spatial resolution for x-coordinates and y-coordinates at eight positions on SBS BT GEM prototype.

The beam spot positions on the SBS BT GEM prototype during the scan are shown on the left plot in Figure 14. These values yield a measured resolution of $\sigma_x \approx 60$ μm for x-strips and of $\sigma_y \approx 68$ μm for y-strips at this location. The right plot in Figure 14 shows the spatial resolution for x and y coordinates at eight different locations on the detector. The resolution shows good uniformity across the detector with an average resolution of 59.53 μm for x-strips and 62.88 μm for y-strips. The x-coordinate corresponds to the top readout layer, with narrower strips leading to lower capacitance and lower noise while the y-coordinate corresponds to the bottom layer, with wider strips and slightly higher capacitance and higher noise. Moreover, the bottom electrodes collect fewer charges than the top strips because of the non-equal charge sharing discussed in paragraph 4.2. These two effects could explain the 5 μm difference between resolutions $\sigma_x$ and $\sigma_y$. These resolutions are consistent with the expected performances for standard Triple-GEM detectors. The slightly higher value of $\sigma_y$ at the location P5 is explained by the presence of 300 μm spacers at this location causing a local degradation of the spatial resolution.

## 5. Conclusion

A 50 × 50 cm² GEM module was designed and prototyped at the University of Virginia for the Back Tracker GEM chambers of the Super Bigbite Spectrometer at Jefferson lab. Two prototype modules were extensively tested using cosmic, radioactive sources and test beam data. These data indicate that the detectors perform as expected meeting the requirements of the Super Bigbite Spectrometer.

**Acknowledgment & Disclaimer**

This work has been accepted for publication in Nuclear Inst. and Methods in Physics Research, A.

We thank Leszek Ropelewski and Eraldo Oliveri at the GDD/RD51 Detector Lab and Rui de Oliveira at the PCB production facility, all at CERN, for their help and technical support with the construction of the detector. We also thank Evaristo Cisbani at INFN Roma (Italy) for the valuable input into the design, characterization and optimization of the detector prototypes. Finally, we acknowledge the funding support of Jefferson Science Associates, LLC under U.S. DOE Contract No. DE-AC05-06OR23177.